\documentclass[%
 aip,
 amsmath,amssymb,
 reprint,%
]{revtex4-1}

\usepackage{graphicx}
\usepackage{dcolumn}
\usepackage{bm}

\usepackage{epsfig}
\usepackage{amsmath}
\usepackage{braket}
\usepackage{physics}

\newcommand{\beq}{\begin{equation}}
\newcommand{\eeq}{\end{equation}}
\newcommand{\bea}{\begin{eqnarray}}
\newcommand{\eea}{\end{eqnarray}}

\begin{document}
\renewcommand{\thefootnote}{\fnsymbol{footnote}}
\title{Guided matter wave inertial sensing in a miniature physics package}

\author{K. D. Nelson}
\author{C. D. Fertig}
\affiliation{Honeywell International Inc., 12001 Highway 55, Plymouth, MN 55441, USA}
\author{P. Hamilton}
\altaffiliation[Now at ]{Dept. of Physics, University of California, Los Angeles, CA 90095, USA}
\author{J. M. Brown}
\altaffiliation[Now at ]{Physical Sciences Inc., Andover, MA 01810, USA}
\author{B. Estey}
\altaffiliation[Now at ]{Honeywell International Inc., Broomfield, CO 80021, USA}
\author{H. M\"uller}
\affiliation{Department of Physics, University of California, Berkeley, CA 94720, USA}
\author{R. L. Compton}
\email{robert.compton3@honeywell.com}
\affiliation{Honeywell International Inc., 12001 Highway 55, Plymouth, MN 55441, USA}

\begin{abstract}
We describe an ultra-compact ($\sim 10$ cm$^3$ physics package) inertial sensor based on atomic matter waves that are guided within an optical lattice during almost the entire interferometer cycle.  We demonstrate large momentum transfer (LMT) of up to 8$\hbar k$ photon momentum with a combination of Bragg pulses and Bloch oscillations with scalability to larger numbers of photons.  Between momentum transfer steps, we maintain the atoms in a co-moving optical lattice waveguide so that the atoms are in free space only during the Bragg pulses.  Our guided matter wave approach paves the way for atomic inertial sensing in dynamic environments in which untrapped atoms would otherwise quickly collide with the walls of a miniature chamber.
\end{abstract}


\maketitle

Inertial sensing is foundational to modern era navigation systems, which tyically rely on a filtered combination of inputs from inertial measurements and the global positioning system (GPS)\cite{Brown:2012}.  With growing concerns around the vulnerability of GNSS to jamming and spoofing \cite{Jafarnia:IJNO2012, C4ADS:2019}, it is desirable to identify inertial sensing technologies that can enable navigation for long periods of time in the absence of GNSS correction.  Atom interferometers hold great promise due to the traceability of inertial sensitivity to stable quantities like the momentum transferred to an atom by a photon \cite{Peters:Metrologia2001, Peters:Nature1999, Durfee:PRL2006}.  Several groups have demonstrated mobile atom interferometers, but even the most recent versions represent initial steps, at best, toward meeting size, weight, and power (SWaP) requirements for most terrestrial and aerospace applications  \cite{Muller:EPJD2009, Wu:Stanford2009, Bodart:APL2010, Menoret:ICSO2010, Schmidt:Gyro2011, Kitching:IEEE2011, Geiger:NatComm2011, McGuinness:APL2012, Hauth:ApplPhysB2013, Rakholia:PRAppl2014, Battelier:SPIE2016, Wu:Optica2017, Wu:arXiv20019}.

\begin{figure}
	\includegraphics[width=8cm]{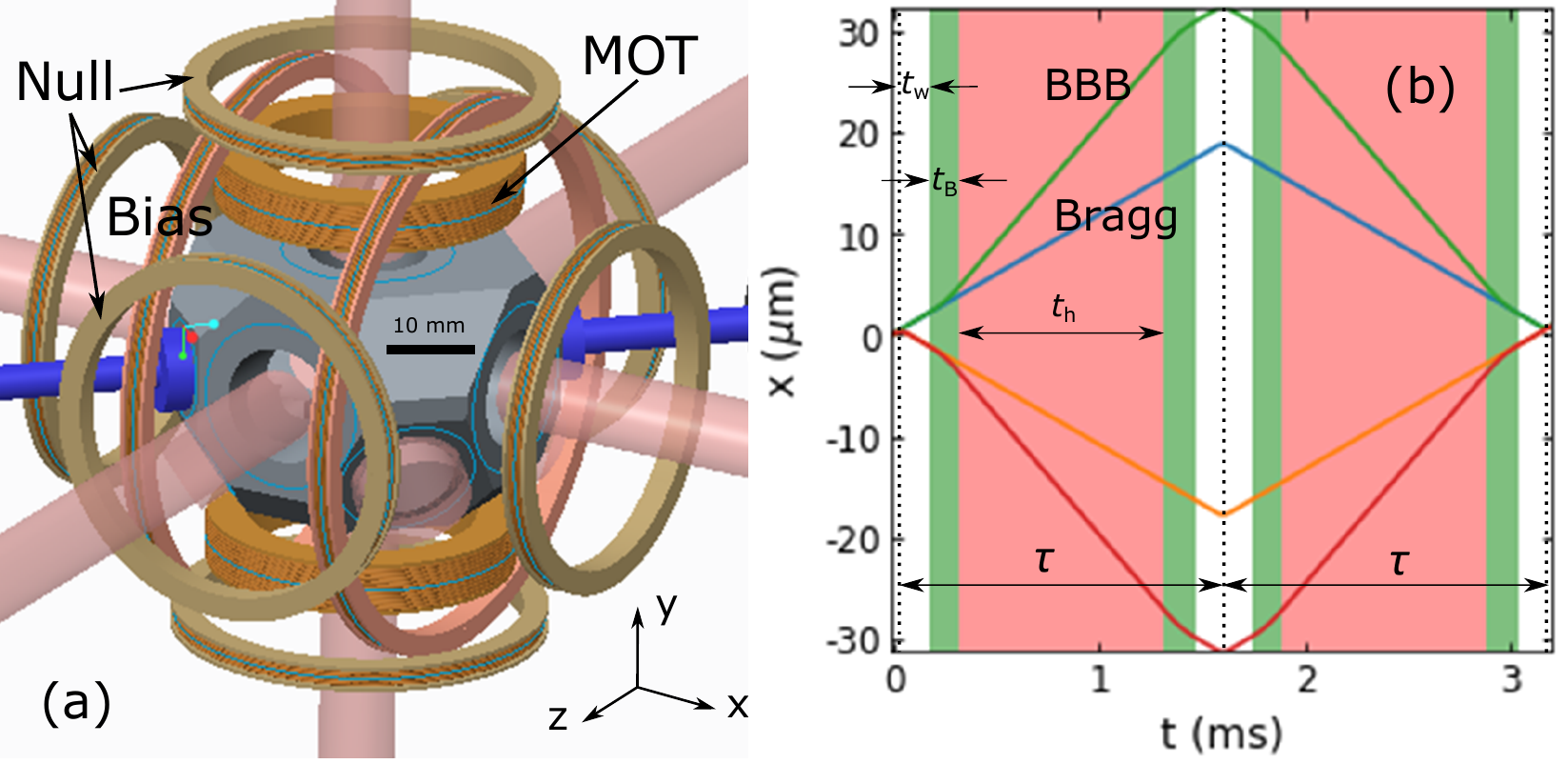}
 \caption{(a) Drawing of miniature $\sim 10$ cm$^3$ physics package, machined from BK7 glass with AR coated windows attached by glass frit process.  Magnetic field coils are wound on non-metallic cores, including MOT coils oriented along the \textit{y}-axis, field nulling coils along all axes, and a bias field coil along the \textit{x}-axis.  Gravity is along the \textit{y}-axis and optical momentum transfer is along the \textit{x}-axis.  Titanium tubes (blue) are attached by glass frit to conduct vacuum and insert Rb metal.  (b) Separation of BBB and Bragg-only interferometer paths as a function of interferometer cycle timing. The dashed vertical lines denote time of Bragg diffraction pulses.  Green shading denotes acceleration due to Bloch oscillations for the BBB path only.  The pink shaded region denotes a co-moving lattice for the BBB path only. The time $t_w$ is a wait time between Bragg and Bloch sections, $t_B$ is the duration of the Bloch acceleration, $t_h$ is the hold time in the lattice in-between BOs, and $\tau$ is the Bragg pulse separation time.}
    	\label{fig:BBBTrajectory}
\end{figure}

Sensor scale factor is proportional to the separation distance between the arms of the interferometer, so the key challenge for miniaturization is to obtain a sufficiently large separation distance within a short interrogation time, before the atoms collide with the wall of the chamber as a result of their own thermal motion, or the acceleration of the platform, a goal that aligns with a desirable sensor bandwidth and update rate, reducing performance requirements for solid state co-sensors \cite{Lautier:APL2014}. One potential path toward greater miniaturization is the use of large momentum transfer (LMT) techniques to increase the rate of separation between the arms, enabling larger scale factor with a shorter interrogation time.  LMT based on high order Bragg transitions is limited by constraints on atom cloud temperature \cite{Giltner:PRL1995, Muller:PRL2008}, while LMT based on consecutive light pulses is limited by the compounding of inefficiencies with each Raman transition \cite{McGuirk:PRL2000}. 

These limitations to LMT interferometry can be overcome by a combination of small momentum transfer atom optics (Raman or Bragg beamsplitter and mirror) followed by Bloch oscillations to accelerate the separation velocity between the two arms of the atom interferometer \cite{Clade:PRL2009, Muller:PRL2009, Charriere:PRA2012}.  In the Bragg-Bloch-Bragg (BBB) scheme, the two arms of the interferometer preserve a common internal state, which greatly reduces deleterious phase shifts due to the AC Stark effect \cite{Muller:PRL2009} , making this approach particularly suitable to inertial sensing \cite{Hamilton:IEEE2014}.  Recently, high contrast fringes have been demonstrated in a BBB configuration with up to 408 $\hbar k$ photon momenta between interferometer arms \cite{Gebbe:arXiv2019},  as well as in an optical cavity assisted interferometer that provides deep lattices at large detunings, enabling extreme robustness against vibration \cite{Xu:Science2019}.  To date, however, there have been no demonstrations of LMT atom interferometry in a miniature package that is comparable in size to state-of-the-art navigation grade inertial sensors.

Here, we implement a BBB interferometer loaded from magneto-optically trapped atoms in a miniature $\sim 10$ cm$^3$ vacuum chamber, demonstrating progress toward robust atomic inertial sensing in small, low-cost packaging.  With large momentum transfer nearly perpendicular to local gravity, we report an initial factor of four increase in scale factor sensitivity for a small projection of gravity along the acceleration sensitive axis.  In doing so, we note that the atoms are supported in a co-moving optical lattice during portions of the interferometer cycle where there is no Bloch acceleration, so that this approach represents a nearly fully guided matter-wave interferometer, in which the atoms are well supported against gravity and other accelerations.  This work therefore represents a milestone on the path toward robust atomic inertial sensing in packaging that is comparable in size, weight, and power consumption to state-of-the-art optical gyros.  We also discuss sensitivity to rotation, which arises as the atoms accelerate along the the gravitational vector, perpendicular to the axis of optical momentum transfer, thereby enclosing an area.

As shown schematically in Fig.~\ref{fig:BBBTrajectory}(a), we implemented a Bragg-Bloch-Bragg (BBB) interferometer \cite{Muller:PRL2009} in a compact vapor cell $\sim 10$ cm$^3$ containing laser cooled $^{87}$Rb atoms supplied by benchtop laser sources.  A master laser was locked to an optical transition near 780.241 nm using saturated absorption spectroscopy.  Cooling light was furnished by a slave laser that was locked at a frequency offset relative to the master using a beat-note between the two lasers.  A second slave laser, also referenced to the master, was locked either 3~GHz red relative to the $F=1 \rightarrow F'$ transition to drive Raman transitions or approximately 12~GHz blue relative to the $F=1 \rightarrow F'$ transition to drive Bragg and Bloch transitions.  Raman beams were delivered with circular polarization, while Bragg and Bloch lightwaves were linearly polarized.  Raman, Bragg, and Bloch $k$-vectors were oriented along the \textit{x}-axis of Fig.~\ref{fig:BBBTrajectory}(a), with a $1/e^2$ beam waist of 3.5 mm.  To generate Raman sidebands, light from the second slave laser was modulated at 6.834 GHz using a fiber-coupled electro-optic modulator (EOM).  To remove the carrier and unwanted sidebands, light from the EOM was split and then directed through a pair of temperature stabilized etalons, each tuned to transmit at only one of the Raman transitions.

The schematic trajectory shown in Fig.~\ref{fig:BBBTrajectory}(b) forms a Mach-Zehnder atom interferometer with three Bragg pulses of duration $t_{\pi}$ that transfer $4\hbar$k momentum and are separated in time by $T$.  In between the Bragg pulses, Bloch oscillation (BO) sections transfer addtional momentum $2 N \hbar k$ to each arm, for a combined total momentum separation between arms of $(4 + 4 N) \hbar$k.  The combination of Bragg and Bloch oscillations has been referred to as a Bragg-Bloch-Bragg (BBB) interferometer \cite{Muller:PRL2009}.  The phase of a BBB atom interferometer can be shown (see Supplemental Material) to be

\beq
	\phi = 4 k g T [ T + N (t_B + t_h )],
\label{eq:PertInt}
\eeq
where we note that $N$ is the number of Bloch oscillations for \textit{each} arm, while $4 N \hbar$k is the additional momentum separation between the two arms, due to Bloch oscillations.

The experimental cycle begins with 210 ms of laser cooling in a magneto-optical trap (MOT).  The MOT coils are switched off, and the MOT field decays over 5 ms to a value less than 100 mG, established by a combination of magnetic field nulling coils and a single layer magnetic shield.  During this 5 ms period, the cooling light is detuned by 10's of MHz to support polarization gradient cooling, resulting in a population $\sim 10^6$ atoms at a temperature of $\sim 5 \mu K$.  Approximately 1/3 of these atoms are distributed into the $\ket{ F=1, m_F=0}$ sublevel by the cycling transition after the repump light is turned off.  Next, a bias field $\sim 100$ mG establishes a quantization axis along the \textit{x}-axis (see Fig.~\ref{fig:BBBTrajectory}(a).  From the magnetically insensitve $\ket{ F=1, m_F=0}$ population, a narrow velocity class is transferred from $\ket{ F=1, m_F=0, p = 2 }$ to $\ket{ F=2, m_F=0, p = 0 }$ by a $150~\mu$s Raman $\pi$ pulse applied along the \textit{x}-axis, while the remaining atoms are subsequently distributed across the seven magnetic sublevels of $\ket{ F=2}$ by a $500~\mu$s repump step.  A second Raman $\pi$ pulse, also of duration $150 \mu$s, now performs a velocity selection from $\ket{ F=2, m_F=0, p = 0 }$ back into $\ket{ F=1, m_F=0, p =2 }$, followed by 1 ms of resonant light to blow away the remaining atoms in $\ket{ F=2 }$, leaving a population of indeterminate size in the $\ket{ F=1, m_F=0 , p = 2 }$ state with a Gaussian velocity distribution having a FWHM $\sim 1 v_{\textit{rec}}$ along the selection axis, corresponding to a 1D temperature $T \approx 60$ nK.  The atoms remain in the $\ket{ F=1, m_F=0 }$ internal state for the entire interferometer sequence.

After a 100 $\mu$s delay for settling of laser locks, the interferometer cycle begins.  The atomic wavefunction is split by a Bragg $\pi/2$ beamsplitter pulse of $60 \mu$s duration, shaped with a Gaussian temporal profile having FWHM $\approx 10~\mu$s and peak power $\approx 15$ mW.  For the beamsplitter pulse, the frequencies of the lattice beams $f_1$ and $f_2$ were equal in the lab frame, placing the atoms in a superposition of $\ket{ p=2}$ and $\ket{ p=-2 }$ momentum states.  The lattice light is turned off for $10~\mu$s and then the intensity in each of \textit{three} beams is linearly ramped to $\approx 8$ mW over $100~\mu$s to form a double optical lattice in which each lattice is co-moving with one of the divided momentum states.  The double lattice was formed by passing one of the lattice beams through an AOM that was being driven at two different frequencies $f_2$ and $f_3$, while the other lattice beam remained at the same frequency $f_1$ that had been used for the Bragg beamsplitter pulse.  The total time between the peak of the Bragg pulse and the beginning of Bloch oscillations is $t_w = 140 \mu$s.  Next, in order to drive Bloch oscillations, frequencies $f_2$ and $f_3$ are ramped through $+8 f_r$ and $-8 f_r$ respectively, where the recoil frequency $f_r \approx 3.77$ kHz for $^{87}$Rb.  The Bloch ramp is linear in frequency, occuring on a timescale of $t_B = 150~\mu$s, resulting in population transfer to the $\mid p = \pm 4 \rangle$ states, giving a total momentum separation of $8~\hbar$k for the two halves of the wavefunction.

The atoms remain in the lattice for a hold time $t_h$ that was varied from 1 $\mu$s to 2 ms before reversing the process.  In reverse, the lattice is ramped back down in frequency  and then in intensity, and finally left off for 10 $\mu$s before driving a Bragg $\pi$ pulse.  The Bragg $\pi$ pulse has the same temporal width as the Bragg $\pi /2$ beamsplitter, but a higher intensity $\approx 25$ mW, and acts as a mirror to swap the momentum states of the two halves of the wavefunction.  The process of raising the lattice and sweeping Bloch oscillations is then repeated, ending with a final Bragg $\pi/2$ pulse.

Following the interferometer cycle, readout occurs by mapping velocity states onto internal states and integrating fluorescence from spectrally distinct states.  A 150 $\mu$s velocity sensitive Raman pulse transfers atoms from $\ket{ F=1, m_F = 0, p=2}$ to $\ket{F=2, m_F=0, p=0}$.  The atoms are allowed to expand ballistically for 1.5 ms, and then the first readout pulse occurs, with fluorescence from the cycling transition collected on a photodetector, measured via a transimpedance amplifier, and integrated for $200 \mu$s to produce a single voltage $V_a$.  The atoms are repumped for $40 \mu$s resulting in the entire population collected in the $\ket{F=1}$ state and then the fluorescence collection is integrated for another $200 \mu$s resulting in a total fluorescence signal $V_b$.  Finally after $1.5$ ms of continuous probe light to blow away remaining cold atoms, the background vapor fluorescence level $V_c$  is measured and integrated for $200 \mu$s.  The fraction of atoms in the $\ket{p=2}$ arm of the interferometer is then taken to be $\lvert \bra{p=2}\ket{\Psi} \rvert^2 =(V_a - V_c)/(V_b - V_c)$.  Not mentioned above are two technical delays of $400-500 \mu$s prior to the $V_b$ and $V_c$ fluorescence collections, to ensure that the phototdetector integrators have been cleared of all prior signals.  

In the second case, all of the same timings were used, but there was no lattice light on during any of the lattice steps, so only a Bragg interferometer was implemented, with a maximum momentum separation of 4 $\hbar$k between halves of the wavefunction.

\begin{figure}
	\includegraphics[width=8cm]{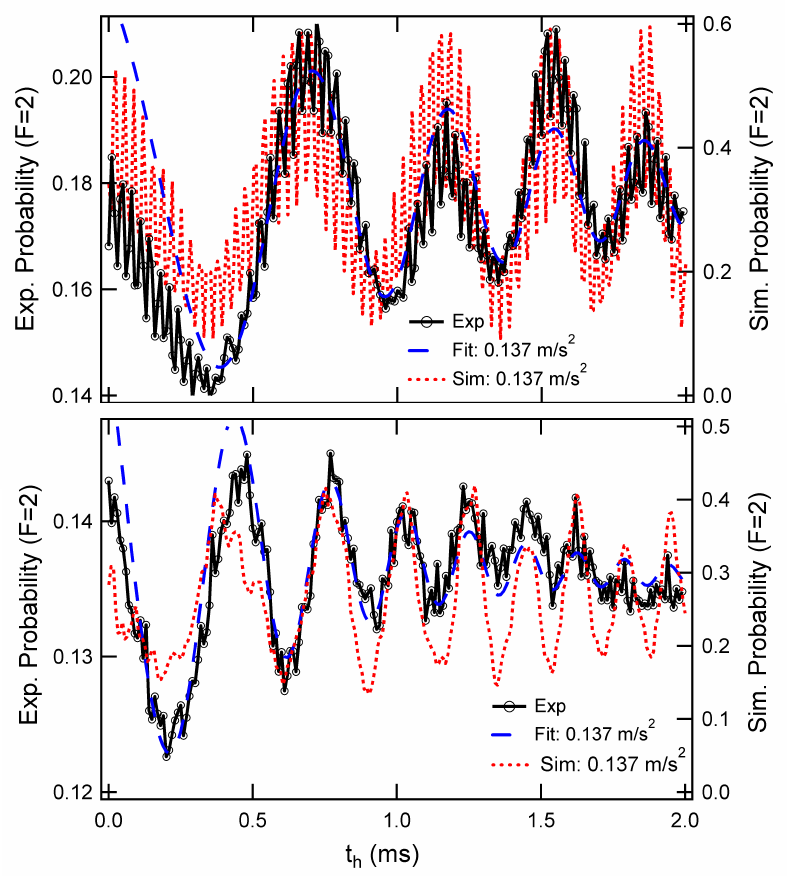}  
 	 \caption{Interferometer fringes for a $4\hbar k$ Bragg interferometer (a) and an $8\hbar k$ BBB interferometer (b), both with a small component of gravity projected onto the acceleration sensitive axis by tilting the sense axis slightly away from horizontal.  Fraction of atoms in $\ket{v=2}$, determined by mapping the velocity state onto internal state $\ket{F=2}$.  Open circles are experimental data, connected by solid lines as a guide to the eye.  Dotted red lines are the results of numerical simulations, and the dashed blue line is a fit to Eq.~\ref{eq:PertInt}.}
    	\label{fig:TimeSeries}
\end{figure}

The interferometer was mounted on an optical table, with the acceleration sense axis (optical axis for lattice light) tilted slightly away from horizontal by adjusting the pneumatic floats, resulting in a small projection of gravity onto the sense axis.  We first implemented the simple 4 $\hbar$k Bragg interferometer, without Bloch oscillations, resulting in the interferometer fringes of Fig.~\ref{fig:TimeSeries}(a), which shows the fraction of atoms in the $\ket{p=2}$ velocity state as a function of holding time $t_h$ at the point of maximum separation velocity.  Open circles denote experimental data obtained from an average over 21 data sets and are connected by solid black lines as a guide to the eye.

The dashed blue lines in Fig.~\ref{fig:TimeSeries} are fits to Eq.~\ref{eq:PertInt}, with $T =2( t_w + t_B)+t_h$, and including phenomenological DC offset, linear, and exponential decay terms.  The fits were restricted to data between $t_h = 500~\mu$s and 2 ms, in order to mitigate the effects of the finite duration $t_B=30~\mu$s of the Bragg pulses.  For Fig.~\ref{fig:TimeSeries}(a), the fit yields $g = 0.1370 \pm 0.0012$ m/s$^2$ for the projection of gravity along the accelerometer sense axis, consistent with an approximate measure of tilt angle based on measured heights of the corners of the optical table relative to the floor.  The fit yields a decay time $\tau_{\text{data}} \approx 1.1$ ms, which can be compared with numerical simulations.  Numerical solutions of the Schr\"odinger equation, based on a Crank-Nicolson solver \cite{Compton:PRA2012}, are shown as dotted red lines, and yield a favorable qualitative fit to the data for the same value of projected $g$.  Monte Carlo treatment of cloud temperature, with $T = 60$ nK, yields a noticeable exponential decay of the simulation data, although with a slightly longer time constant than observed in the data, suggesting that decoherence of the interferometer signal is dominated by external factors, especially platform vibration.  Single photon scattering due to the 12 GHz detuned Bragg beams is negligible.

A rapid oscillation as a function of holding time $t_h$, with a period of approximately 30 $\mu$s, is superposed on Bragg fringes of Fig.~\ref{fig:TimeSeries}(a).  These oscillations are narrow interference fringes whose period corresponds approximately to the inverse Rabi frequency of the Bragg mirror transition.  These narrow interference fringes are well reproduced in numerical simulations and are due to undesired transitions between $\ket{p=\pm2}$ and $\ket{p=0}$, occuring because the energy separation $\approx 15$ kHz x $h$ between the states is easily accessed by the comparable Rabi frequency for the Bragg pulses.

Fig.~\ref{fig:TimeSeries}(b) shows the results of the full BBB interferometer, with $N$ = 1, meaning that one Bloch oscillation has been added to each arm of the interferometer, for a total of 8$\hbar$k peak momentum separation for the two paths, including equal contributions from Bragg and Bloch processes.  The fit to Eq.~\ref{eq:PertInt}, again restricted to $500 \mu$s $<t_h<2$ ms, yields $g=0.1366 \pm 0.0010$ m/s$^2$, consistent with the previous Bragg result, and a significantly shorter decay time $\tau_{\text{BBB}} \approx 580 \mu$s.  The shorter decay time is likely related to the overall reduced contrast as compared to the simpler Bragg-only case, but cannot be attributed to a greater level of single photon scattering, which is only $\sim 1 \%$ over an entire 2 ms cycle, even though the atoms spend all but $\sim 100 \mu$s in the 12 GHz blue detuned lattice.  Note that two lattices are always present - a blue detuned lattice that is co-moving with the atoms and traps them at intensity nodes, minimizing scattering, plus a lattice that is moving in a direction opposite to the atoms, and therefore affects the atoms with an RMS average scattering intensity but does not drive Bloch transitions.  Simulations do not adequately capture the overall loss of contrast for the BBB interferometer, but otherwise yield a reasonable qualitative fit to the chirped sinusoid. 

\begin{figure}
	\includegraphics[width=8cm]{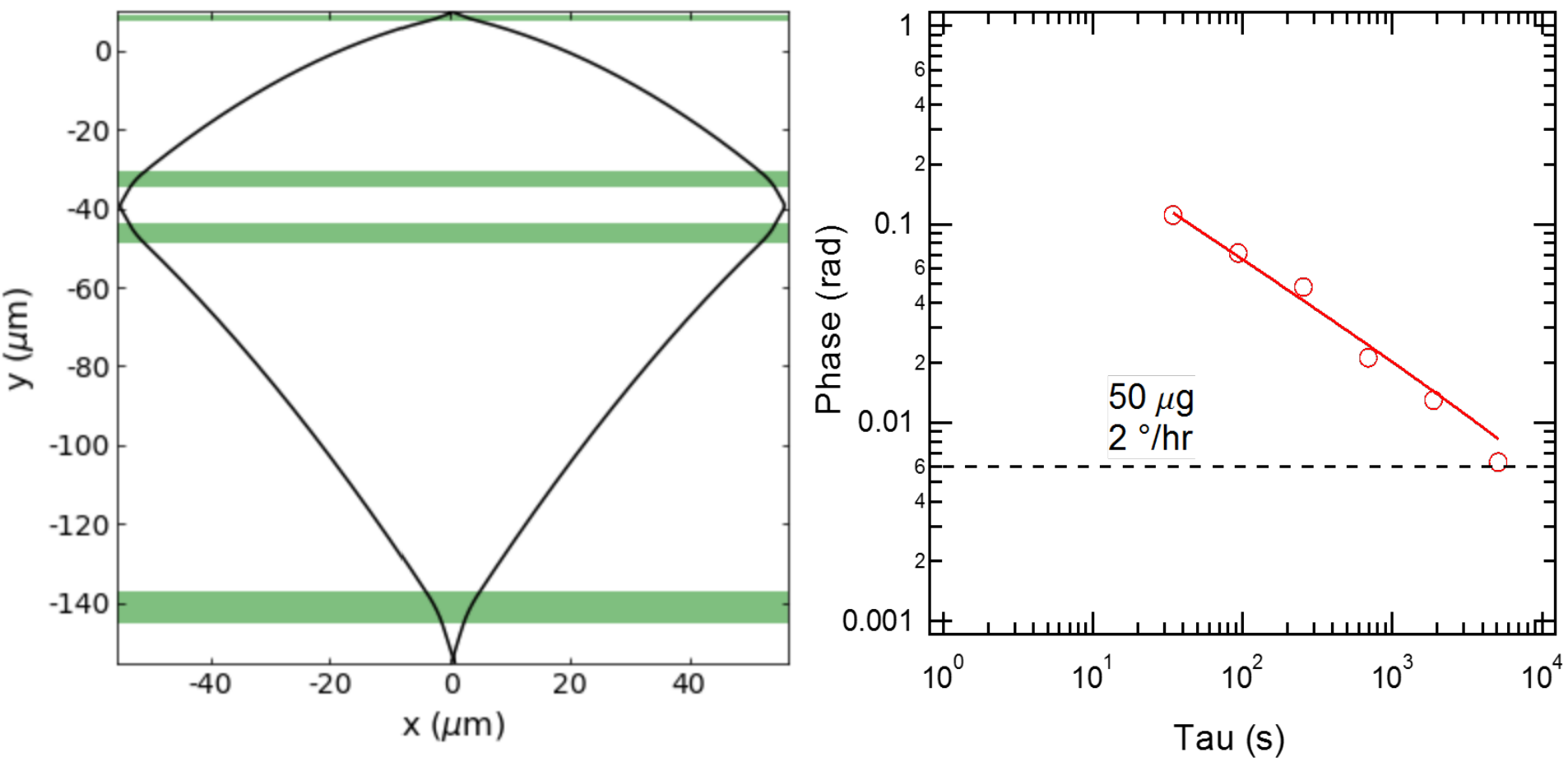} 
 	 \caption{(a) Calculated trajectory for matter waves falling in the $y$-direction while being separated by lightwaves with a $k$ vector along $x$.  The area enclosing path provides sensitivity to rotations.  A second optical lattice (not shown) applied along the $y$-direction would enable scalability to navigation grade gyroscope performance.  (b) Allan variance of the phase shift for the Bragg Bloch interferometer.  Application of inertial scale factors to these data indicates stability to at least 50 $\mu$g and 2 $^\circ$/hr at $10^4$ s of integration.}
    	\label{fig:GyroPath}
\end{figure}

We note that the horizontal orientation of the accelerometer sense axis means that the gravitational acceleration vector is very nearly perpendicular to the $k$-vector of optical momentum transfer.  Because of this, the trajectories of the waveguide paths move downward as they separate, so that the interferometer path encloses an area, as in Fig.~\ref{fig:GyroPath}, which causes the interferometer phase to be sensitive to rotations about the \textit{z}-axis of Fig.~\ref{fig:BBBTrajectory}(a) in addition to accelerations along the \textit{x}-axis.  With only a single interferometer, we are unable to simultaneously measure finite acceleration and rotation, but can nevertheless investigate interferometer responses in a lab setting where inertial inputs are well controlled.  For the inferred trajectory Fig.~\ref{fig:GyroPath}, the sensitivity at $10^4$ seconds of integration time is $2^\circ$/hr., which supports sensitivity to Earth rate $\approx 15^\circ$/hr.  However, in the present laboratory setup, we were unable to rotate the sense axis with respect to Earth's rotation axis in order to distinguish the contribution to the phase shift due to Earth's rotation from other systematic contributions.  

We have demonstrated an accelerometer based on a nearly fully guided matter wave interferometer.  Using the Bragg-Bloch-Bragg approach, we have demonstrated a scale-factor enhancement of nearly a factor of four relative to a conventional Raman interferometer with 2 $\hbar k$ momentum separation between the arms.  Because the atoms are supported by an optical lattice against gravity and other accelerations, the accelerometer has potential to perform robustly for missions requiring high dynamic range.  The BBB interferometer contrast was significantly degraded due to decoherence of the wavefunction, likely arising from laser intensity noise, but this can be mitigated in future implementations.  

The sensitivity for this approach will likely be driven by the need for ultra-cold atoms in order to obtain high efficiency Bragg transitions required for the atom optics (beamsplitter and mirror pulses).  Raman velocity selection, used here to obtain a sub-recoil population, discards a large fraction of the Maxwell-Boltzmann distribution obtained from optical molasses.  Advanced cooling techniques, such as free space Raman cooling \cite{Kasevich:PRL1992, Boyer:PRA2004} or Raman sideband cooling \cite{Hamann:PRL1998, Kerman:PRL2000}, would be necessary to in order to boost SNR to competitive levels.
\\

{\bf SUPPLEMENTAL MATERIAL}

Supplemental material, below, derives analytical expressions for the phase evolution of the Bragg-Bloch-Bragg interferometer, following both perturbative and non-perturbative approaches.  Both approaches result in the expression of Eq.~\ref{eq:PertInt}.
\\

{\bf ACKNOWLEDGMENTS}

We acknowledge helpful discussions with J. Dorr and C. Hoyt.
We acknowledge funding under the DARPA Chip-Scale Atomic Navigator (C-SCAN) program under AMRDEC contract W31P4Q-13-C-0092.  The views, opinions, and/or findings expressed are those of the authors and should not be interpreted as representing the official views or policies of the Department of Defense or the U.S. Government.
The data that support the findings of this study are available within the article and its supplementary material.
\\


{\bf SUPPLEMENTAL MATERIAL}

The interferometer shown in Fig.~\ref{fig:BBBTrajectory} is a Mach-Zehnder atom interferometer with three Bragg pulses of duration $t_{\pi}$ that transfer $4\hbar$k momentum and are separated in time by $T$.  In between the Bragg pulses, Bloch oscillation (BO) sections transfer addtional momentum $2 N \hbar k$ to each arm, for a combined total momentum separation between arms of $(4 + 4 N) \hbar$k.

The combination of Bragg and Bloch oscillations has been referred to as a Bragg-Bloch-Bragg (BBB) interferometer \cite{Muller:PRL2009}.  The paths shown in Fig.~\ref{fig:BBBTrajectory} are best defined by specifying the acceleration as a function of time.  We assume that the inteferometer is initiated with an upward momentum of $2\hbar$k, and that the first Bragg laser pulse, occuring at $t=0$, transfers $-4\hbar$k.  We will assume that the Bragg pulses have negligible duration, an assumption that works for $T\gg t_\pi$.  Accelerations for the upper $a_u$ and lower $a_l$ paths as a function of time $t$ are

\begin{align}
& a_u = -g - 4 v_r \delta (t-T) \notag + 4 v_r \delta(t-2T) \\
	& + a_B \begin{cases}
	~1, \text{if}\ & (t_w < t  < t_w + t_B) \\
	 & \| (2T-t_w-t_B < t < 2T-t_w) \\
	-1 , \text{if}\ & (T-t_w- t_B < t < T- t_w ) \\
	& \| (T+t_w < t < T+t_w+t_B), \\
	\end{cases} \notag \\
& a_l = -g -4 v_r \delta(t) + 4 v_r \delta (t-T) \notag \\
	& + a_B \begin{cases}
	-1, \text{if}\ & (t_w < t  < t_w + t_B) \\
	& \| (2T-t_w-t_B < t < 2T-t_w) \\
	~1 , \text{if}\ & (T-t_w- t_B < t < T- t_w ) \\
	& \| (T+t_w < t < T+t_w+t_B), \\
	\end{cases}
\label{eq:accelerations}
\end{align}
where $g$ is the acceleration of free fall, $v_r = \hbar k/m$ is the single-photon recoil velocity for an atom of mass $m$ and laser wavenumber $k$, $\delta(t)$ is the Dirac delta function, and $a_B$ is the magnitude of the acceleration during the Bloch section, in addition to $g$.  All quantities in Eq.~\ref{eq:accelerations} are measured in the rest frame of the apparatus.  From the accelerations, the velocity and position of the atoms are calculated by integration.

The phase of atom interferometers due to a constant (spatially) gravitational acceleration $g$ can be calculated perturbatively.  For this purpose, we first evaluate the trajectories Eq.~\ref{eq:accelerations} for $g=0$.  Such an interferometer has zero phase difference between the arms, as can be shown by direct calculation.  With the unperturbed trajectories as a starting point, we then estimate the phase for $g\neq 0$ by integrating the potential $m g z(t)$ over these unperturbed trajectories:

\begin{align}
	\phi 	&= \frac{1}{\hbar} \int{ m g} [ x_u(t, g=0) - x_l( t, g=0)] dt \notag \\
		&= 4 k g T [ T + N (t_B + t_h )],
\label{eq:PertInt}
\end{align}
where we again note that $N$ is the number of Bloch oscillations for \textit{each} arm, while $4 N \hbar$k is the additional momentum separation between the two arms, due to Bloch oscillations.  Calculating the trajectories and the integral is straight-forward but tedious and will be omitted here.


In the non-perturbative approach, we break down the total phase $\Delta \phi$ as
\beq
	\Delta \phi = \Delta \phi_F + \Delta \phi_L,
\label{eq:PhiSum}
\eeq

where the first term is the free evolution phase difference $\Delta \phi_F$, obtained by integrating the Lagrangian along the closed path of the interferometer, and the second term is the laser phase $\Delta \phi_L$. Whenever the atom absorbs a photon, the local phase of the photon is added to the atomic matter wave packet, and subtracted for stimulated emission.  To evaluate the phase, we note that $\Delta \phi_F = 0$, as in the usual Mach-Zehnder atom interferometer.  We then calculate the laser phase due to the Bragg diffractions, and finally the laser phase due to the Bloch oscillations.

\textit{Free Evolution Phase}:  The contributions to the free evolution phase

\beq
	\Delta \phi_F = \frac{1}{\hbar} \oint{ L(t) dt }
\label{eq:Lagrange}
\eeq

that arise when all light is off can be calculated by standard methods. This section considers those parts that arise during Bloch oscillations. During Bloch oscillations, the atoms are in a potential

\beq
	V(z) = -V_0 \cos^2{ [k (z-a t^2/2] + m \bm{g} \cdot \bm{z} }
\label{eq:LatticePot}
\eeq

By choosing the minimum of the periodic potential to be nonzero, we have assumed that the lattice is red detuned.  The atoms are assumed to follow the lattice. With constant Bloch acceleration $a$, the coordinates of the particle are thus

\beq
	z(t) = z_0 + v_0 t + \frac{a t^2}{2},~~~~\dot{z}(t) = v_0 + a t.
\label{eq:coords}
\eeq

The phase accumulated during the time the lattice is accelerating is given by
\beq
	\phi_B = \phi^g_B + \phi^T_B + \phi^V_B + \phi^{osc}_B
\eeq
where
\beq
	\phi^g_B = -\frac{gm}{\hbar} \int{z(t) dt}
\eeq
is the phase due to the gravitation potential,
\beq
	\phi^T_B = -\frac{m}{2 \hbar} \int{(v_0+a t)^2 dt}
\eeq
the phase due to the kinetic energy,
\beq
	\phi^V_B = \begin{cases}
	0 & \text{blue detuned} \\
	\abs{V_0} t_B/ \hbar & \text{red detuned} \\
	\end{cases}
\eeq
the phase due to the lattice potential, and
\beq
	\phi^{osc}_B = -\frac{1}{2}\omega^{osc} t_B
\eeq
the phase from the zero-point energy of the lattice harmonic oscillator.

Assuming that $\phi^V_B$ and $\phi^{osc}_B$ are the same for each Bloch oscillation section, we can add the free evolution phase accumulated in between Bloch oscillations and find that
\beq
	\Delta \phi_F = 0.
\label{eq:phiFree}
\eeq
This important result is arrived at in a laborious but otherwise straightforward calculation.

\textit{Laser phase}  The laser phase from Bragg diffraction is calculated in the usual way:
\beq
	2k(x_u(0) - x_u(T)) - x_l(T) + x_l(2T) = 4 k g T^2,
\label{eq:phiBragg}
\eeq
recalling that the path separation due to the Bragg pulse reflects $4k\hbar$k of momentum transfer.  

To evaluate the laser phase during Bloch oscillations, we look at one of the Bloch oscillation sections, which we assume starts with atoms in each arm at a relative velocity $v_0$ at $t=0$.  The total velocity change for each arm is $\pm 2N v_r$ due to optical lattices accelerating both upward and downward.  One lattice beam remains constant in frequency, while two additional lattice beams are swept with a frequency difference that is started at $\pm \delta \omega = 2 k v_0$ and ends at $\pm \delta \omega = 2 k (v_0 + 2 N v_r)$, so
\beq
	\pm \delta \omega(t) = 2 k v_0 + 4 k N v_r t/ t_B.
\eeq

The center of mass of a particle in each arm moves according to
\beq
	\pm x(t) = x_0 + v_0 t + N v_r t^2 / t_B.
\eeq

Momentum transfer occurs when the particles reach the end of the Brillouin zone, which happens for the first time when the particle has been accelerated by $v_r$, at $t = t_B/ (2N)$, and then in intervals of $t_B/N$, i.e., at $t_n = t_B (2n-1)/ 2N$.  At these times, the location is
\beq
	\pm x_n = x_0 + v_0 t_B \frac{2n-1}{2N} + v_r t_B \frac{(2n-1)^2}{4N}.
\eeq
Assuming the acceleration is positive, the corresponding laser phase is $2k \sum^N_{n=1}{x_n}$, where
\begin{align}
	\sum\limits_{n=1}^N x_n &= \sum\limits_{n=1}^N \left ( x_0 + v_0 t_B \frac{2n-1}{2N} + \frac{v_r t_B (2n-1)^2}{4N} \right) \notag \\
					&= N x_0 + \frac{1}{2} N t_B v_0 + t_B v_r \frac{4 N^2 - 1}{12}.
\end{align}

This laser phase now has to be summed over the four Bloch oscillation sections in the interferometer. To simplify this, note that the $t_B v_r$ term is independent of initial velocity or position, and thus common to all four Bloch sections. It cancels out. To sum up the $N k t_B v_0$ term, we note that the initial velocities $v_0$ take the values

\begin{align}
	v_1 &= 2 v_r - g t_w \notag \\
	v_2 &= (2 + 2 N) v_r - g (t_w + t_B + t_f), \notag \\
	v_3 &= 2 v_r - g( T + t_w), \notag \\
	v_r &= (2 + 2N) v_r - g (T+t_w + t_B + t_h)
\end{align}

Thus, we have
\begin{align}
	& k N t_B ( v_1 -v_2 -v_3 + v_4) \notag \\
	& = k N t_B \biggl[  2 v_r - g t_h -2 v_r -2 N v_r \notag \\
	& + g (t_w + t_B + t_h)  -  2 v_r + g ( T + t_w) \notag \\
	& + (2+2N) v_r -g (t+t_w + t_B + t_h)  \biggr] \notag \\
	& = 0
\end{align}
A nonzero contribution will be made by the $x_0$ term.  We denote the four different initial $x$-values by $x_{1-4}$. We need to calculate
\beq
	2k N (x_1 - x_2 - x_3 + x_4)
\eeq
where $x_2 -x_1 + g T (t_h + t_B) = x_4 - x_3$ and thus
\beq
	\pm 2 k N (x_1 - x_2 - x_3 + x_4) = \pm 2 N k g T (t_h + t_B)
\eeq

Since the free evolution phase vanishes, Eq.~\ref{eq:phiFree}, we add the laser phase from Bragg diffraction Eq.~\ref{eq:phiBragg} and Bloch oscillation, including contributions to both arms, and obtain
\beq
	\phi = 4k g T [T + N (t_f + t_B)]
\eeq
which is equivalent to the perturbative result.

\end{document}